# MRI Super-Resolution with Ensemble Learning and Complementary Priors

Qing Lyu, Hongming Shan, and Ge Wang, *Fellow, IEEE*

*Abstract*—Magnetic resonance imaging (MRI) is a widely used medical imaging modality. However, due to the limitations in hardware, scan time, and throughput, it is often clinically challenging to obtain high-quality MR images. The super-resolution approach is potentially promising to improve MR image quality without any hardware upgrade. In this paper, we propose an ensemble learning and deep learning framework for MR image super-resolution. In our study, we first enlarged low resolution images using 5 commonly used super-resolution algorithms and obtained differentially enlarged image datasets with complementary priors. Then, a generative adversarial network (GAN) is trained with each dataset to generate super-resolution MR images. Finally, a convolutional neural network is used for ensemble learning that synergizes the outputs of GANs into the final MR super-resolution images. According to our results, the ensemble learning results outcome any one of GAN outputs. Compared with some state-of-the-art deep learning-based super-resolution methods, our approach is advantageous in suppressing artifacts and keeping more image details.

## I. INTRODUCTION

Magnetic resonance imaging (MRI) is one of the most commonly used medical imaging modalities over the world. It noninvasively depicts structural and functional features inside a patient in rich contrasts. Compared with other medical imaging modalities such as computed tomography (CT) and nuclear imaging, MRI does not involve ionizing radiation. Furthermore, MRI plays a dominating role in neurological imaging especially brain research.

However, MRI suffers from major physical limitations in background magnetic field, gradient fields, imaging speed, and signal-to-noise ratio. Hence, the pixel size of a clinical MR image is often constrained to an order of millimeters. To obtain higher resolution MR images, either a more complicated MRI system with stronger background magnetic field or a longer scan time needs to be used, which will dramatically increase expenses and is rarely clinically applicable. To address this problem, the super-resolution (SR) approach holds a great promise that needs to change neither hardware nor scanning protocol [1], [2].

SR has been a hot topic in the computational vision field over the past decades. A large number of methods were proposed to improve image resolution retrospectively. Roughly speaking, super-resolution algorithms can be divided into the two categories: model-based and learning-based. Interpolation algorithms [3], [4], like bilinear, bicubic, and nearest neighbor interpolation techniques, are representatives of simple model-based approaches, which can be directly used to enlarge images. Wiener filtering and iterative deblurring algorithms utilize knowledge on the system point spread function (PSF) to recover image resolution [5], [6], and are also considered as model-based algorithms. Dictionary learning-based super-resolution techniques [7]-[9] are examples of learning-based algorithms.

Recently, with the rapid evolution of machine learning, especially deep learning, deep neural networks have become popular among SR studies. In this context, diverse neural network architectures were designed and tested. Successful architectures include convolutional neural networks (CNNs) [10]-[16], generative adversarial networks (GANs) [17]-[20], residual neural networks (ResNets) [21]-[24], and recurrent neural networks (RNNs) [25]-[27]. Specifically, for MRI SR studies, models based on CNN, ResNet, and GAN were reported [28]-[32]. Although there were a plenty of decent SR results in the literature, bothersome artifacts or distortions were often observed in the SR results produced by these methods. Therefore, the challenge remains for achieving superior SR on MR images. In other words, MRI SR research must deliver images with faithful details and little artifacts in order to be clinically relevant. Since the existing methods cannot perfectly satisfy such a requirement, it is necessary to develop new SR methods based on all encouraging results.

Among important deep learning strategies, ensemble deep learning is a meta-method for further enhancement of deep learning results [33], [34]. By combining multiple models, ensemble learning is expected to produce synergistic results than the results from individual models. Merits of ensemble learning were successfully demonstrated to improve image quality [35], [36]. In this MRI SR study, by combining multiple GAN models trained on different image prior datasets, we demonstrate better MRI SR image quality than that associated with any single model. In this paper, we describe a

*(Corresponding author: Ge Wang)*
All authors are with the Department of Biomedical Engineering, Rensselaer Polytechnic Institute, Troy, NY, 12180, USA (email: lyuq@rpi.edu, shanh@rpi.edu, wangg6@rpi.edu).

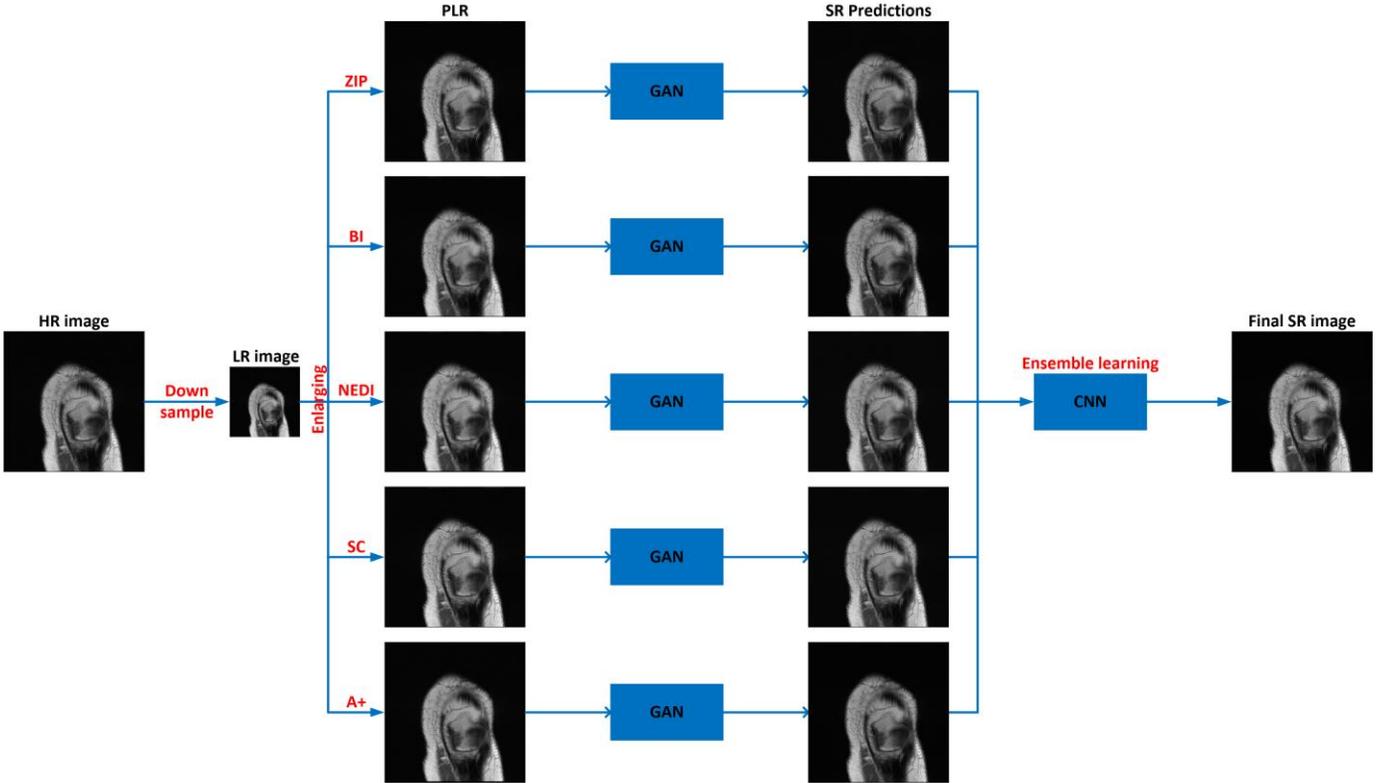

Fig. 1. Overall workflow of the SR method proposed in this paper.

convolutional neural network to integrate diverse SR results from individual models. Our method produces better image quality than state-of-the-art SR models, being capable of detailing textures and avoiding artifacts.

## II. METHODOLOGY

Assuming that $I_{LR} \in \mathbb{R}^{p \times q}$ is a low-resolution (LR) MR image of the size $p \times q$ and $I_{HR} \in \mathbb{R}^{m \times n}$ the corresponding high-resolution (HR) MR image of the size $m \times n$. The relationship between them can be expressed as

$$I_{LR} = f(I_{HR}) \qquad (1)$$

where $f: I_{HR} \in \mathbb{R}^{m \times n} \to I_{LR} \in \mathbb{R}^{p \times q}$ denotes the down-sampling/blurring process that creates a LR counterpart from a HR image. The overall SR process is to implement an approximate inverse function $g \approx f^{-1}$

$$g \approx f^{-1}: I_{HR} \approx g(I_{LR}) \qquad (2)$$

As an ill-posed problem, it is impossible to find this inverse function exactly. Instead, priors can be used to acquire an approximation of $f^{-1}$. Multiple types of priors are used in this study to approximate $f^{-1}$ as closely to the ideal function as possible.

Fig. 1 presents the proposed overall workflow of SR method. In our scheme, we first down-sample original HR images and obtain LR counterparts, and then use five different model-based SR algorithms to process LR images and obtain five image datasets, each of which corresponds to one of the five SR algorithms. These image datasets are separately fed into the corresponding GAN for training and then to produce five SR counterparts. The final step is to use a CNN to do the ensemble learning and obtain the final MRI SR result.

### A. Down-sampling

Differing from commonly used down-sampling strategy that down samples data in the image domain, we chose to down-sample data in the frequency domain. The original HR images were converted into the frequency domain via the 2D Fourier transform. Only the central 25% of the data points were kept in the *k*-space, while all peripheral data points were zeroed out. Then, the cropped dataset was converted back into an image through the 2D inverse Fourier transform as a LR image. Using this down-sampling method, we obtained LR images with 2-fold resolution degradation, which means $p = m/2$ and $q = n/2$.

### B. Processing

Before LR images are fed into the neural network, we adopt existing SR algorithms to enlarge a LR image to the size of the target HR image. This processing step was also used in some SR studies like VDSR [13] and SRCNN [11]. With such a processing step, the quality of enlarged images can be somehow improved so that it is easier for neural networks to extract image features and obtain better prediction results. Moreover, we can establish datasets with complementary priors through using multiple SR algorithms in this processing step. In this paper, in order to distinguish enlarged images with the latter neural network results, we call enlarged images "processed LR images (PLR)" rather than SR images. In this work, 5 classic SR algorithms were used to enlarge LR images and obtain the corresponding PLRs, which are zero interpolation filling (ZIP) [37], bicubic interpolation (BI) [3], new edge directed interpolation (NEDI) [4], sparse coding (SC) [7], and adjusted anchored neighborhood regression (A+) [9]. ZIP is a simplest super-resolution method that adds zeros in the peripheral region in the frequency domain. It is widely used to enlarge MR images. BI and NEDI are two improved interpolation-based algorithms, while SC and A+ are two dictionary-learning-based algorithms. In the A+ algorithm, 1,024 dictionary atoms were used, which were obtained from a training dataset of 0.5 million natural images. Similarly, the SC method also used a dictionary contained 1,024 atoms. Using these different SR algorithms, different enlarged images were created, each with its unique priors. Next, we used these enlarged images as the inputs to train the corresponding GAN models separately. Each enlarging/deblurring process can be expressed as

$$I_{PLR,i} = h_i(I_{LR}) \quad (3)$$

where $I_{PLR,i} \in \mathbb{R}^{m \times n}$ is a deblurred/enlarged image. $h_i: I_{LR} \in \mathbb{R}^{p \times q} \to I_{PLR,i} \in \mathbb{R}^{m \times n}$ denotes the processing process. As there were 5 SR algorithms used in this processing step, $i = 1,2,\ldots,5$.

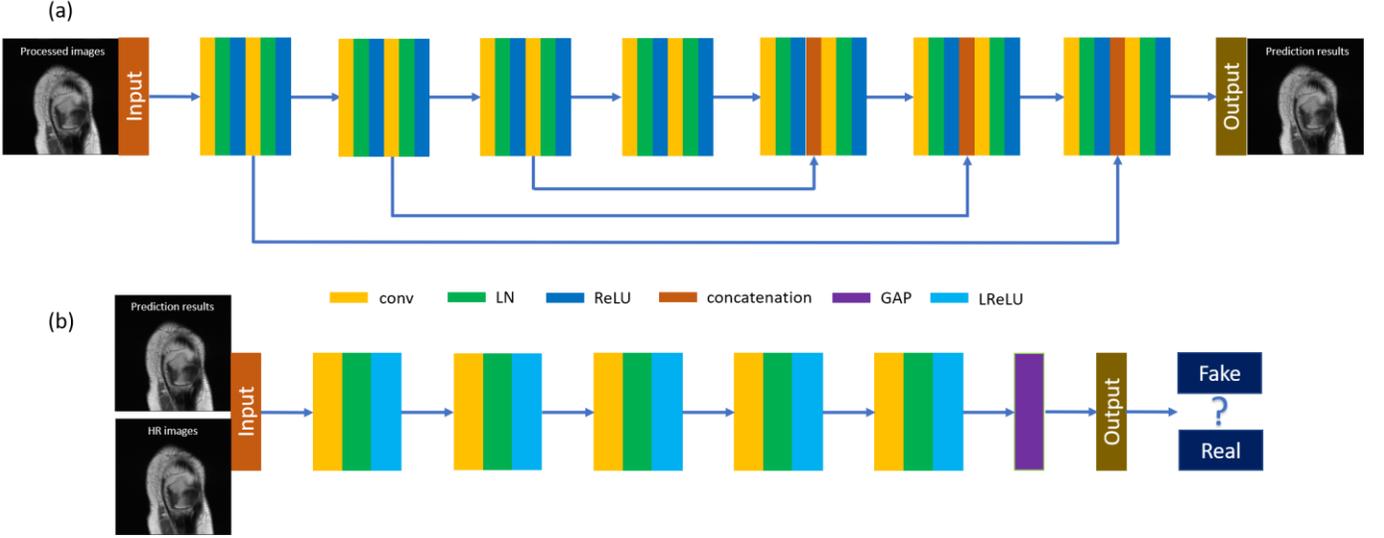

Fig. 2. The architecture of the proposed GAN. (a) The generator and (b) the discriminator.

*C. GAN model*

The GAN model with Wasserstein distance and gradient penalty (WGAN-GP) [38] is selected in this study. The generator used in this study is a modified version of the encoder-decoder that was originally used for CT denoising [39]. The architectures of the generator and the discriminator are shown in Fig. 2. For the generator, it contains seven blocks with three skip connections. Each block involves convolution, layer normalization (LN), and ReLU activation operations sequentially. As mentioned in [38], batch normalization cannot be used in the WGAP-GP model, instead, we utilize layer normalization. With skip connections, feature maps in the previous blocks are concatenated to the feature maps in the latter blocks. Because of the structural similarity between LR and HR images, adding such skip connections can greatly boost the training process and keep the generated images structurally similar to the input LR images. Similarly, the discriminator consists of five blocks followed by a global average pooling (GAP) layer. Each block includes convolution, layer normalization, and leaky-ReLU (LReLU). Note that we replace the fully-connected layer with a GAP layer to reduce the amount of parameters in the neural network and facilitate the training process [44]. Each block in the discriminator performs convolution, layer normalization and ReLU activation function similar to what the block in the generator does. The details of the generator and the discriminator are listed in Table I and II.

The objective function of the generator is as follows:

$$\min_{\theta_G} \mathcal{L}_G = \mathcal{L}_{adv} + \lambda_1 \mathcal{L}_{gra} + \lambda_2 \mathcal{L}_{mse} + \lambda_3 \mathcal{L}_{per} \quad (4)$$

$$\mathcal{L}_{adv} = -\mathbb{E}_{I_{PLR}}[D(G(I_{PLR}))] \quad (5)$$

$$\mathcal{L}_{gra} = \mathbb{E}_{(I_{PLR}, I_{HR})} \|\nabla(G(I_{PLR})) - \nabla(I_{HR})\|^2 \quad (6)$$

$$\mathcal{L}_{per} = \mathbb{E}_{(I_{PLR}, I_{HR})} \|\emptyset(G(I_{PLR})) - \emptyset(I_{HR})\|^2 \quad (7)$$

where $\theta_G$ stands for the generator parameters. $\mathcal{L}_{adv}$ is the adversarial loss. $\mathcal{L}_{gra}$ is the gradient loss, which is the L2 norm difference between gradients of the generated image and its ground truth. $\nabla$ stands for the gradient calculated in both x and y axis. $\mathcal{L}_{mse}$ and $\mathcal{L}_{per}$ are the mean squared error (MSE) and the perceptual loss respectively. MSE reflects the image content similarity between generated output images and processed input images at the pixel level, while the perceptual loss shows the similarity in a high-level feature space. We use the pretrained VGG19 model to calculate the perceptual loss [39], [40], and $\emptyset$ represents the feature map of the 16[th] convolutional layer in the VGG19 model. As the mean-squared-error loss tends to blur images, the gradient loss is introduced to penalize any blurring and enhance the edge. The objective function of the discriminator is the standard WGAN-GP equation shown in [38]. Overall, this GAN SR process is then expressed as

$$I_{SR,i} = G(I_{PLR,i}) = \varphi(I_{PLR,i}, \theta_{G,i}) \qquad (8)$$

where $I_{SR,i} \in \mathbb{R}^{m \times n}$ is a output image from generator in GAN, $\varphi: I_{PLR,i} \in \mathbb{R}^{m \times n} \to I_{SR,i} \in \mathbb{R}^{m \times n}$ denotes a GAN deblurring process for an SR image, $\theta_{G,i}$ represents the parameters of a GAN model. As there are 5 SR algorithms used in this step, $i = 1,2,\ldots,5$.

TABLE I
STRUCTURAL DETAILS OF THE GENERATOR, WHERE n1s1 MEANS 1 CONVOLUTION KERNELS WITH STRIDE OF 1.

| Block1 | Block2 | Block3 | Block4 | Block5 | Block6 | Block7 |
|---|---|---|---|---|---|---|
| n32s1 | n64s1 | n128s1 | n256s1 | n128s1 | n64s1 | n32s1 |
| n32s1 | n64s1 | n128s1 | n256s1 | n128s1 | n64s1 | n1s1 |

TABLE II
STRUCTURAL DETAILS OF THE DISCRIMINATOR, WHERE n1s1 MEANS 1 CONVOLUTION KERNELS WITH STRIDE OF 1.

| Block1 | Block2 | Block3 | Block4 | Block5 |
|---|---|---|---|---|
| n64s2 | n128s2 | n256s2 | n512s1 | n1s1 |

*D. Ensemble learning*

Ensemble learning is to synergize diverse information sources for the best results. In this study, SR images individually obtained can be integrated via ensemble learning to improve SR MR image quality. Specifically, the final SR results are derived from different types of GAN SR predictions. In the processing step, we utilized the five SR algorithms to process LR images and obtain five sets of enlarged/deblurred images. Then, we train five GAN models (with the same structure) separately corresponding to each of these five algorithms and then obtain five sets of prediction results. After that, a CNN model is used to integrate all these images through ensemble learning. The structure of the CNN is the same as the generator in the proposed GAN. Images from each type of prediction results are concatenated before input to the CNN for training, validation, and testing. The mean-absolute-error (MAE) is used in the loss function during the training process. The ensemble learning process can be expressed as

$$I_{SR} = \phi(\sum I_{SR,i}, \Theta) \qquad (9)$$

where $I_{SR} \in \mathbb{R}^{m \times n}$ denotes ensemble learning results which are also our SR outcomes, $\phi: I_{SR,i} \in \mathbb{R}^{m \times n} \to I_{SR} \in \mathbb{R}^{m \times n}$ denotes the ensemble learning process, $\Theta$ represents the parameters of the CNN model.

III. RESULTS

Our ensemble learning SR approach was streamlined and evaluated for MRI SR. It produced SR images with more structural details, less artifacts and distortion.

In the following, we will show our processed images, GAN predictions, and ensemble learning results, and compare our SR results with state-of-the-art methods. To evaluate image quality, the metrics we used include structural similarity (SSIM) and peak signal-to-noise ratio (PSNR).

*A. Experimental implementation and neural network convergence*

The NYU fastMRI dataset was used in this study [41]. All images were reconstructed from proton density weighted knee scans under 1.5 or 3 Tesla. The size of original HR images is $320 \times 320$. Totally, 159 patient scans with 5,744 slices were used for training (80%) and validation (20%), and additional 43 patient scans with 1,500 slices for testing. Before training, each processed image was decomposed into forty-nine $80 \times 80$ patches. During the training process of the network models, weights in the objective function were selected to be 0.1, 0.1, and 1 for $\lambda_1$, $\lambda_2$, and $\lambda_3$ respectively via the grid searching strategy. The training continued for 50 epochs with the learning rate of $2 \times 10^{-5}$ and batch size of 64 for each GAN model. During the training process, to visualize the convergence of the involved models, the values of MSE and perceptual loss are recorded and showed in Fig. 3. For ensemble learning, the CNN was trained over 80 epochs with the learning rate of $2 \times 10^{-5}$ and batch size of 4. The experiment was conducted in Tensorflow on a single GTX 1080Ti GPU.

According to the MSE and perceptual loss curves in Fig. 3, all the models converge well during the training process since all the curves become gradually steady with the number of training epochs. Compared with the MSE curves, it can be found that the curve of A+ is lower than the other four curves, which shows that the A+ based GAN model can achieve the best SR image with the highest similarity at the pixel level. In the high-level space, as extracted by the perceptual loss, both the A+ and ZIP curves show the lowest values while the NEDI curve is the highest. However, all the curves are pretty close to each other after 40 epochs.

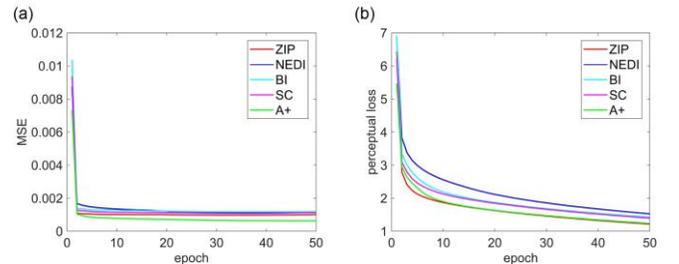

Fig. 3. Comparisons between the involved model types during the training process in terms of (a) MSE and (b) perceptual loss.

*B. Processed results*

It is shown in Fig. 4-6 that PLR images had smoother edges and clearer shapes compared with the corresponding LR images. As different SR algorithms were used in the processing step, PLR images obtained from these SR algorithms were somehow different. As far as the edges in Fig. 5(a) are concerned, it can be found that all the PLR images give clearer edges than the original LR image. Among all the 5 processed images, ZIP and A+ images are closer to the HR image than the other three image types. However, these PLR images also introduced some artifacts and distortions. Some edges in the NEDI PLR image were slightly distorted, with ring or streak artifacts in the ZIP and SC PLR images. In Fig. 5(b), for each PLR image the texture details became smoother while holes in the peripheral region were more distinguishable compared with the LR image. Despite some artifacts introduced, the SC image contains the clearest hole shapes. Each algorithm had its own strengths and weaknesses. After the processing step, we obtained diverse enlarged/deblurred images, some of which present edges well while the others of which preserve details satisfactorily.

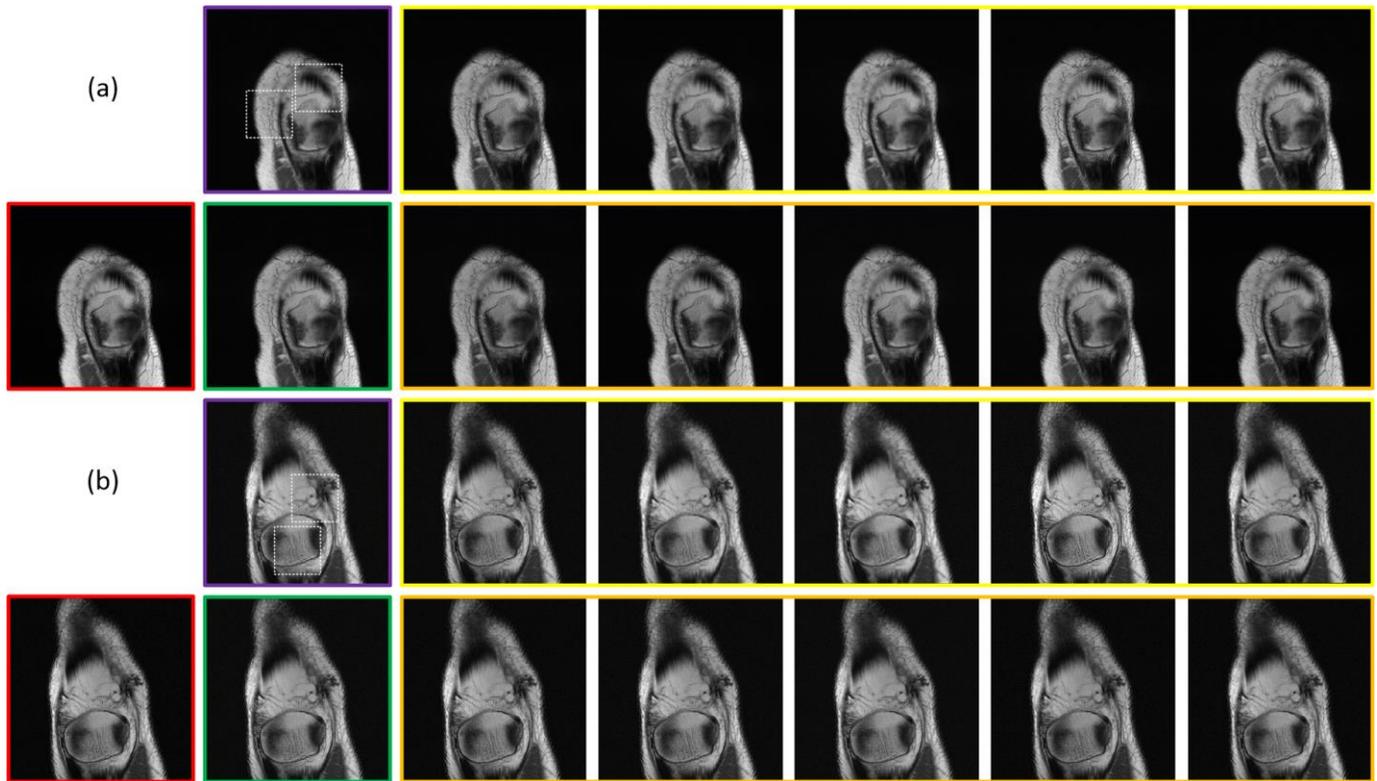

Fig. 4. Input, intermediate and output images in the SR workflow in (a) Cases 1 and (b) 2, where processed results are in the yellow boxes, LR images are in the purple boxes, HR images in the green boxes, ensemble learning results in the red boxes, and GAN predictions in the orange boxes. For the processed results and GAN predictions from left to right are ZIP, BI, NEDI, SC, A+. White dotted boxes are zoomed in region shown in Fig. 5 and 6.

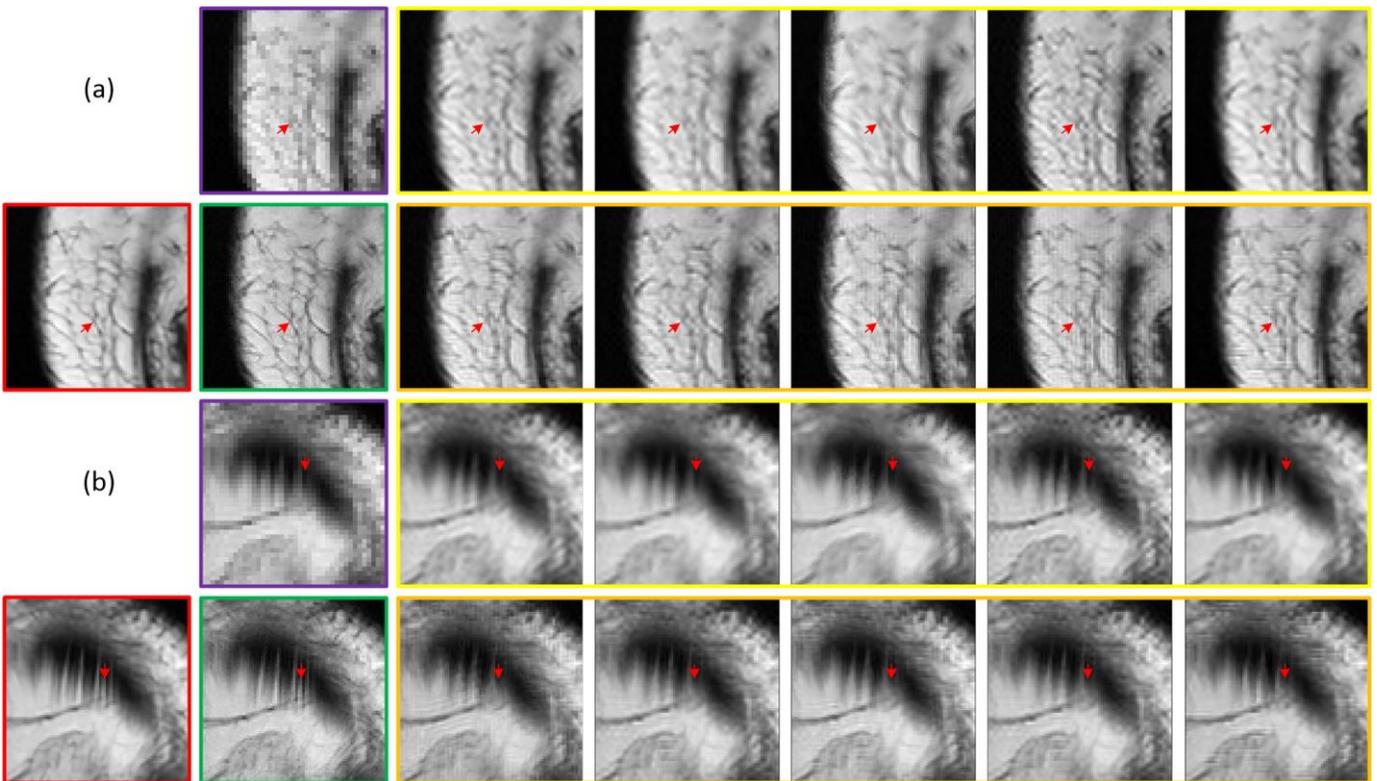

Fig. 5. Zoomed in results of regions marked by white dotted boxes in Cases 1, where processed results are in the yellow boxes, LR images are in the purple boxes, HR images in the green boxes, ensemble learning results in the red boxes, and GAN predictions in the orange boxes. For the processed results and GAN predictions from left to right are ZIP, BI, NEDI, SC, A+.

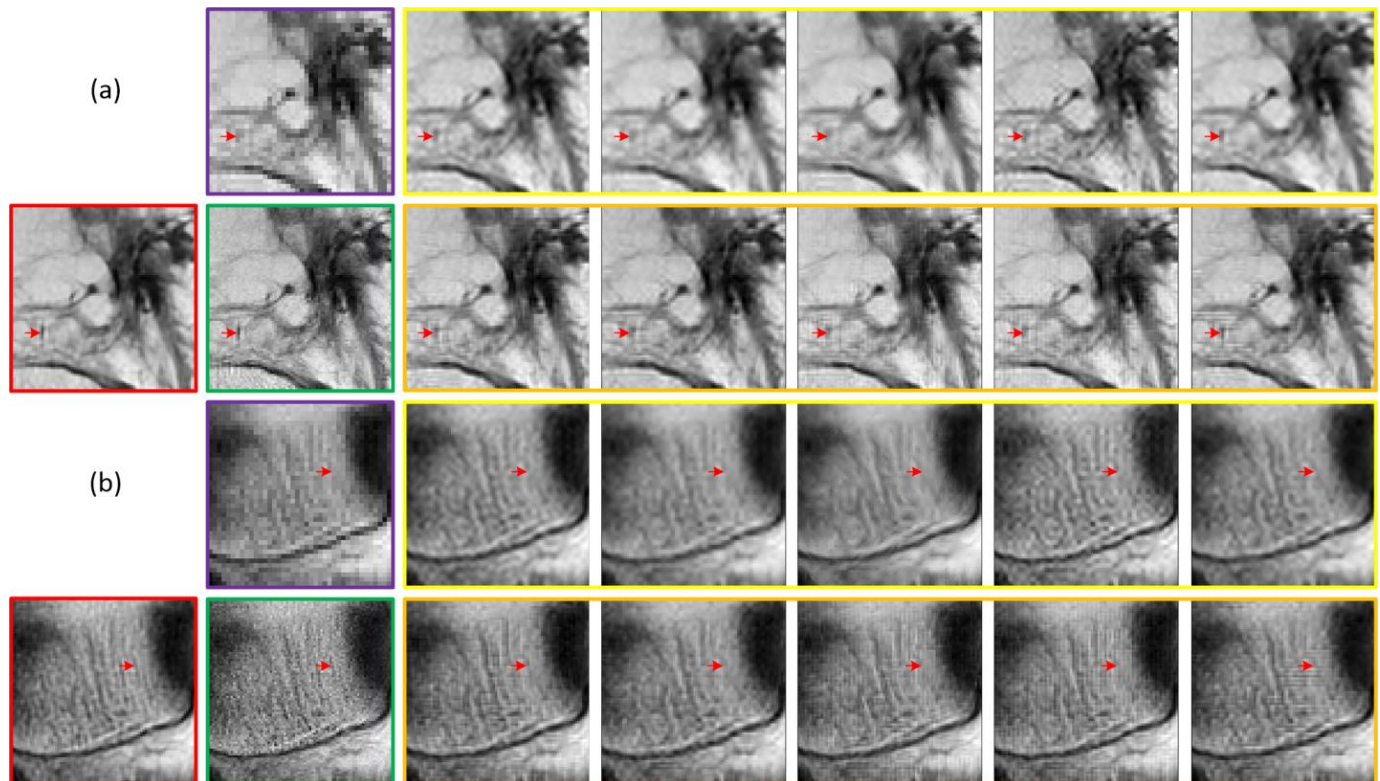

Fig. 6. Zoomed in results of regions marked by white dotted boxes in Cases 2, where processed results are in the yellow boxes, LR images are in the purple boxes, HR images in the green boxes, ensemble learning results in the red boxes, and GAN predictions in the orange boxes. For the processed results and GAN predictions from left to right are ZIP, BI, NEDI, SC, A+.

TABLE III
QUANTITATIVE EVALUATION OF THE PROCESSED IMAGES, GAN PREDICTIONS AND ENSEMBLE LEARNING RESULTS IN FIG. 5 AND 6.

|  |  | Preprocessed results | | | | | GAN predictions | | | | | Ensemble learning |
| --- | --- | --- | --- | --- | --- | --- | --- | --- | --- | --- | --- | --- |
|  |  | ZIP | BI | NEDI | SC | A+ | ZIP | BI | NEDI | SC | A+ |  |
| Figure 4(a) | SSIM | 0.86 | 0.81 | 0.84 | 0.80 | 0.86 | 0.88 | 0.88 | 0.87 | 0.87 | 0.87 | 0.95 |
|  | PSNR | 25.82 | 25.92 | 23.53 | 24.22 | 24.52 | 26.67 | 28.10 | 26.07 | 27.71 | 27.41 | 27.61 |
| Figure 4(b) | SSIM | 0.86 | 0.82 | 0.84 | 0.80 | 0.86 | 0.88 | 0.88 | 0.86 | 0.87 | 0.87 | 0.95 |
|  | PSNR | 28.05 | 26.02 | 26.69 | 25.86 | 28.36 | 28.52 | 28.03 | 26.79 | 28.04 | 27.43 | 32.67 |
| Figure 5(a) | SSIM | 0.85 | 0.79 | 0.82 | 0.77 | 0.85 | 0.86 | 0.86 | 0.85 | 0.86 | 0.86 | 0.92 |
|  | PSNR | 25.23 | 23.49 | 24.96 | 23.92 | 25.95 | 26.90 | 26.73 | 25.74 | 26.34 | 27.15 | 29.44 |
| Figure 5(b) | SSIM | 0.71 | 0.62 | 0.66 | 0.61 | 0.71 | 0.70 | 0.71 | 0.68 | 0.69 | 0.71 | 0.82 |
|  | PSNR | 24.20 | 22.57 | 23.65 | 21.87 | 24.28 | 24.25 | 24.55 | 24.10 | 24.22 | 24.37 | 25.98 |

C. *Super-resolution results through GAN*

After training the GAN models separately, five types of GAN SR prediction results were obtained, as shown in each purple box in in Fig. 3. Compared with the PLR results, the image quality of GAN SR predictions was improved. Generally, they included more details, clearer edges, and less blurs, which are much closer to the original HL images than the PLR versions. For GAN SR predictions, edges in Fig. 5(a) become much clearer while textures in Fig. 6(b) are less blurred compared to the corresponding edges and textures in PLR images. However, just as shown in the PLR images, some artifacts and distortions are also presented in the GAN SR predictions. As the GANs were trained on different training datasets, when GAN SR predictions are compared, differences among them are clearly detectable for the same ground truth. For example, when comparing edges pointed by the red arrow in Fig. 5(a), it can be found that shapes of edges in the GAN SR predictions were slightly different. Another example was shown in Fig. 6(a), shapes of the hole pointed by the red arrows were moderately different from each other. These shape deformations can be called distortions. The quality of the prediction images was quantified in Table III. It can be seen in the table that despite the image quality seems visually improved in those SR prediction results, the PSNR and SSIM values of the GAN predictions were just close to or slightly higher than the PLR results. The appearance of artifacts and distortions in the SR predictions contributed to compromise the PSNR and SSIM values. After training, the GAN SR predictions produced numerous tiny artifacts and distortions. For example, the SSIM value of the ZIP GAN prediction was even lower than the corresponding PLR image. When the two images are compared, despite textures in the GAN SR prediction image were clearer and closer to the HR image, there were streak artifacts produced. As a result, the SSIM value was decreased after training.

### D. Super-resolution results through ensemble learning

Compared with GAN SR predictions, the most distinct characteristic of the ensemble learning results is that artifacts and noise were greatly reduced as shown in Fig. 4-6. Also, the ensemble learning results showed richer textural details than any GAN SR predictions, for example, when comparing the regions pointed by the red arrows in Fig. 5(b), it can be found that the middle peak can be well presented after ensemble learning. In addition, ensemble learning essentially avoided edge distortions, such as the edges pointed by the red arrows in Fig. 5(a) were very close to the ground truth. Quantitatively, the ensemble learning results achieved the highest PSNR and SSIM values among all the results, indicating that ensemble learning greatly improved the image quality.

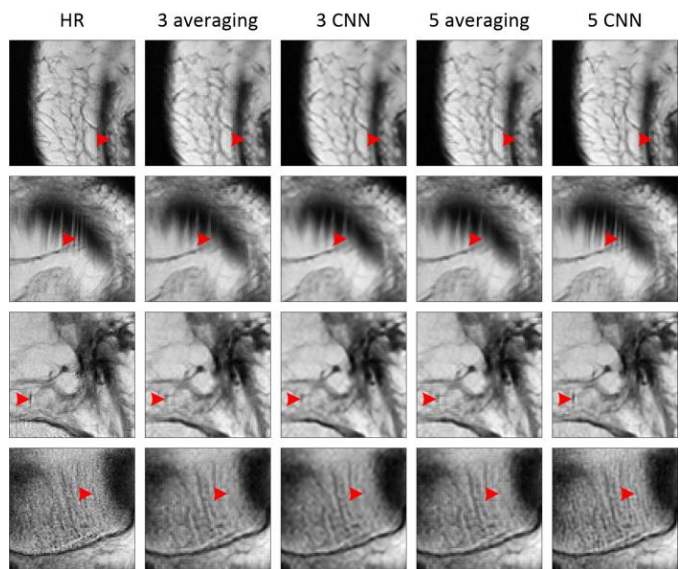

Fig. 7. Ensemble learning results in Cases 1-4, where *k* averaging means that the result is simple average of *k* GAN prediction results, and *k* CNN means that ensemble learning is done through the CNN combining *k* prediction results.

TABLE IV
QUANTITATIVE EVALUATION OF THE RESULTS IN FIG. 7.

|  |  | 3 average | 3 CNN | 5 average | 5 CNN |
|---|---|---|---|---|---|
| Case1 | SSIM | 0.90 | 0.91 | 0.90 | 0.95 |
|  | PSNR | 26.20 | 28.24 | 26.45 | 27.61 |
| Case2 | SSIM | 0.89 | 0.91 | 0.90 | 0.95 |
|  | PSNR | 28.38 | 28.90 | 28.72 | 32.67 |
| Case3 | SSIM | 0.86 | 0.88 | 0.87 | 0.92 |
|  | PSNR | 25.86 | 26.23 | 25.97 | 29.44 |
| Case4 | SSIM | 0.69 | 0.70 | 0.71 | 0.82 |
|  | PSNR | 23.75 | 24.20 | 23.95 | 25.98 |

To investigate how much improvement can be made by the ensemble learning, we further compared our results from ensemble learning with different methods. Instead combining 5 types of GAN SR predictions for ensemble learning, only three types of GAN SR predictions were used. Only GAN SR predictions trained from ZIP, BI and NEDI PLR images were used for the ensemble learning, denoted as 3 CNN. Moreover, simple averaging algorithm was used to replace the CNN model. According to the amount of GAN SR predictions used, they are referred as 3 averaging and 5 averaging. Representative results are presented in Fig. 7 and Table IV. It can be found that the combination of 3 types of GAN SR predictions yielded less PSNR and SSIM values than the integration of 5 types of GAN SR predictions. Importantly, there were more artifacts noticeable in the combination of the 3 types of ensemble learning results, as pointed by the red arrows in Fig. 7. All these results demonstrate that ensemble learning results can be further improved by combining more GAN SR predictions. When comparing the result from simple averaging followed by CNN-based integration, the use of CNN for ensemble learning can achieve better results, showing that optimal integration of diverse SR style would improve image quality further.

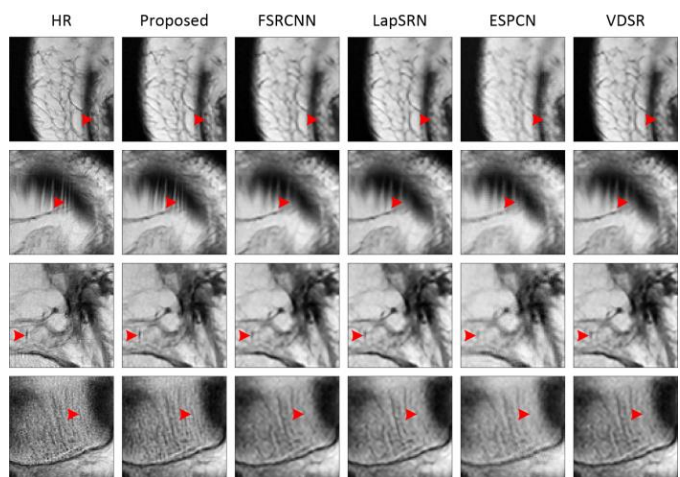

Fig. 8. Comparison of the proposed method and state-of-the-art methods.

TABLE V
QUANTITATIVE EVALUATION OF THE RESULTS IN FIG. 8.

|  |  | FSRCNN | LapSRN | ESPCN | VDSR | Ours |
|---|---|---|---|---|---|---|
| Case1 | SSIM | 0.88 | 0.88 | 0.81 | 0.89 | 0.95 |
|  | PSNR | 25.90 | 25.68 | 21.14 | 26.96 | 27.61 |
| Case2 | SSIM | 0.88 | 0.88 | 0.86 | 0.89 | 0.95 |
|  | PSNR | 28.33 | 28.63 | 24.51 | 28.25 | 32.67 |
| Case3 | SSIM | 0.86 | 0.86 | 0.83 | 0.88 | 0.92 |
|  | PSNR | 25.58 | 25.22 | 21.57 | 26.12 | 29.44 |
| Case4 | SSIM | 0.70 | 0.71 | 0.66 | 0.71 | 0.82 |
|  | PSNR | 24.04 | 23.94 | 21.06 | 24.11 | 25.98 |

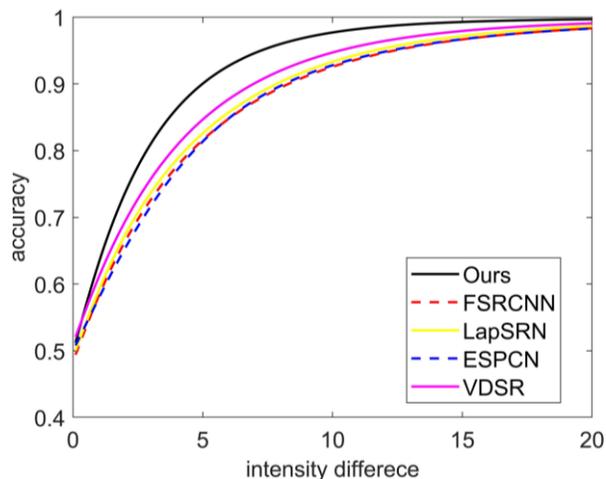

Fig. 9. Pixel intensity accuracy of the proposed method and state-of-the-art methods.

TABLE VI
COMPARISON OF OUR METHOD AND OTHER STATE-OF-THE-ART METHODS (MEAN±STD).

|      | FSRCNN | LapSRN | ESPCN | VDSR | Ours |
|------|--------|--------|-------|------|------|
| SSIM | 0.87±0.02 | 0.88±0.02 | 0.83±0.03 | 0.88±0.02 | 0.92±0.01 |
| PSNR | 28.87±2.33 | 28.80±2.38 | 24.93±3.30 | 28.99±2.25 | 31.35±2.15 |

*E. Comparison with other deep learning-based methods*

Fig. 8-9 and Table V-VI compare our results with the counterparts obtained using other state-of-the-art SR methods. These well-known benchmark SR methods include FRSCNN [12], LapSRN [16], ESPCN [15], and VDSR [13]. Encouragingly, our method outperformed these methods as evidenced by the resultant SR images with higher PSNR and SSIM values. More importantly, our results faithfully present more details and less noise in HR images. For example, in the region pointed by the red arrows in the first two rows of Fig. 8, texture details are clearer and much closer to the ground truth while these textures are blurry in the images produced using the other methods. Also, our results greatly reduced artifacts and distortions. As pointed by the red arrows in the bottom two rows of Fig. 8, the shape of holes and fissure in our images were closer to the ground truth than the competitive counterparts. Fig. 9 compares the pixel intensity accuracies between our result and competing results. Here, we define the accuracy as the percentage of pixels with intensity close to the corresponding pixels in HR images.

$$accuracy = \frac{num_{accu}}{num_{total}} \quad (10)$$

where $num_{accu}$ is the number of accurate pixels and $num_{total}$ is the number of all pixels, and the *x*-axis is the intensity difference threshold. When the intensity difference between a deblurred pixel and its corresponding pixel in the HR image is within this threshold, this pixel is counted as an accurate pixel. For easy comparison, we quantized the all intensity values into the range from 0 to 255. Similar to the receiver operating characteristic (ROC) curve, it is clear in Fig. 9 that the closer a curve is to the left top corner, the more desirable this curve. Compared to the other methods, the curve associated with our proposed ensemble learning method is the closest to the left top corner, indicating that our method produces the results with most pixels close to the ground truth. When the threshold is 5, our method achieved an accuracy of over 90%, which is significantly higher than any one of the other methods.

IV. DISCUSSIONS

Based on the results described above, it can be seen that through the ensemble learning superior MRI SR images can be achieved. In comparison with other state-of-the-art methods, our method produced impressive results without significant artifacts, distortions, and blurry details, giving the highest PSNR and SSIM values.

In the processing step, 5 different SR algorithms were used to enlarge LR images. As a result, 5 types of enlarged PLR image datasets were obtained, each with its unique image priors. From Fig. 4-6, it can be found that each SR algorithm can somehow closely preserve some part of image features. For example, ZIP, BC, and NEDI algorithms are good at presenting edges with clear and smooth shapes, SC and A+ PLR images show texture details. By combining all these SR algorithms together, we can create PLR datasets that are with complementary image priors and finally obtain good SR results that are very close to the ground truth after ensemble learning. Fig. 7 shows the necessity of combining all priors. When only ZIP and interpolation-based methods were used, ensemble learning results would not well present texture details. For example, the middle peak in the second row in Fig. 7 and the fissure in the fourth row in the figure cannot be clearly shown for the 3 CNN method. These fissure and middle peak features are evidently shown in the A+ GAN predictions (shown in Fig. 5(b)), which are not used for the 3 CNN method. Only when all priors are combined in the ensemble learning, we can obtain results with good texture details.

Since introduced in 2014 [42], GAN has been extensively used in many applications due to its compelling performance originated from the adversarial training mechanism. For SR MRI, despite substantial improvements the GAN model tends to induce artifacts and distortions in prediction images, which may due to the nature of GAN and the prevalence of background noise. On one hand, during the training process of a GAN model, the generator is trained to learn the true distribution of training data. Compared with CNN-based supervised training, the training process of GAN is not strongly supervised as predictions are controlled indirectly through a discriminator, not directly anchored with labels. Owing to this mechanism, it is not easy to train the model so that it can perfectly learn the true distribution. As a result, the generator can hardly produce a fake result that is exactly the same as a real result. As shown in Fig. 3, the GAN suffer from edge distortions and ring-like or streak-like artifacts. On the other hand, during the MR scanning process, noise is inevitably recorded [43], by which clinical MR images are greatly influenced. Noise hampers the training process of GAN so that it cannot be perfectly trained. This phenomenon is very evident when noise is prevalent that features cannot be easily extracted. In Fig. 6(b), the subtle textures in the middle region are greatly influenced by noise. As a result, there are substantial artifacts in each of the GAN SR predictions. Fortunately, these issues can be addressed by ensemble learning. By combining multiple GAN SR predictions, the size of the dataset was enlarged. Through a CNN-based integrator, the robustness of the whole SR framework can be strengthened, effectively suppressing artifacts and distortions produced in each GAN SR prediction.

Ensemble learning synergistically combines all GAN SR predictions together. Different from the simple average algorithm, the CNN-based integrator learns to perform ensemble learning in a data-driven fashion and is task-specific. Roughly speaking, the CNN can be assumed as a black box. Through the CNN model, pixels in each input image is assigned for a weight for summation and nonlinear activation. During the training process, the weight of each pixel in each type of GAN SR prediction can be learned according to high-level features and contexts. This data-driven and context-based mechanism guarantees that the CNN-based ensemble learning method outperforms the commonly used averaging method.

Since CNN can approximate a wide class of functions, it is understandable that artifacts and distortions shown in GAN SR predictions can be greatly reduced through CNN-based ensemble learning. As discussed already, ensemble learning enlarges the amount of data with diverse features and increase robustness accordingly. On the other hand, because most artifacts and distortions in each GAN SR prediction were derived from processed images that were obtained using different approaches, these artifacts and distortions were unlikely be the same or very coherent. Through ensemble learning, these artifacts and distortions will be either effectively canceled out or greatly reduced. This ability of alleviating artifacts and distortions is a main advantage of our method compared with other state-of-the-art SR methods.

Another clear advantage of our proposed method is that our results preserve details well. Assuming all subtle features are in HR images, and these features are blurred or weakened in LR images. After the processing step, in each processed image, there are some features strengthened and other features further weakened. For example, in Case 1 in Fig. 3, edges were strengthened through ZIP, and texture was strengthened by A+. Once most features are strengthened by at least one of the processing algorithms, it is in principle possible to collect a relatively complete feature set and recover all details through ensemble learning. As a result, the final SR images after ensemble learning are with clear details, outperforming individual model-based methods.

Compared to other model-based methods, our method is featured by ensemble learning that combines multiple neural network models together, but it demands a higher computational cost.

## V. CONCLUSION

We have proposed an ensemble learning based SR method to improve the image quality of MR images. We have first used 5 existing SR algorithms to enlarge LR images and obtained PLR images with complementary priors. Then, a GAN framework has been used to produce SR prediction results based on different PLR images. Next, we can obtain superior SR images through a CNN-based ensemble learning. The SR images obtained after ensemble learning is better than any individual PLR images or GAN SR prediction results. Compared with other state-of-the-art methods, our method enjoys advantages of minimal artifacts, little distortions, insignificant noise, and rich details.

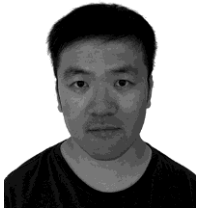

**Qing Lyu** received his Bachelor of Science degree and Master of Science degree in the department of biomedical engineering in Shanghai Jiao Tong University in 2013 and 2016 respectively. In 2017, he joined Dr. Ge Wang's lab. His main research interests include the MRI postprocessing and signal optimization.

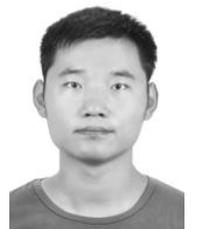

**Hongming Shan** received the B.S. degree from Shandong University of Technology, China, in 2011, and the Ph.D. degree from Fudan University, China, in 2017. He is currently a Postdoctoral Scholar with the Rensselaer Polytechnic Institute, Troy, New York, USA. His research interests include machine/deep learning, computer vision, dimension reduction, and biomedical imaging.

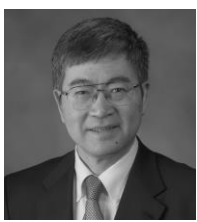

**Ge Wang** is the Clark & Crossan Endowed Chair Professor and Director of the Biomedical Imaging Center, Rensselaer Polytechnic Institute, Troy, NY, USA. He published the first spiral/helical cone-beam/multi-slice CT algorithm in 1991 and has since then systematically contributed over 100 papers to theory, algorithms, artifact reduction and biomedical applications in this important area of CT research. Currently, there are 100+ million medical CT scans yearly with a majority in the spiral/helical cone-beam/multi-slice mode. Dr. Wang's group developed interior tomography theory and algorithms to solve the long-standing "interior problem" for high-fidelity local reconstruction, enabling omni-tomography ("all-in-one") with CT-MRI as an example. He initiated the area of bioluminescence tomography with a significant impact on biophotonics. He has published 460+ journal publications, covering imaging-related diverse topics at depth and receiving a high number of citations and academic awards. His results were featured in Nature, Science, PNAS, and various news media. In 2016, he published the first perspective on deep-learning-based tomographic imaging as the new direction of machine learning and the Nature Machine Intelligence paper on the superiority of deep learning over iterative reconstruction. His team has been continuously well-funded by federal agencies and major imaging companies, actively translating deep learning techniques into imaging products. His research interests include x-ray CT, MRI, optical tomography, multimodality fusion, and machine learning. He is the Lead Guest Editor for five IEEE Transactions on Medical Imaging Special Issues, Founding Editor-in-Chief of International Journal of Biomedical Imaging, Board Member of IEEE Access, and Associate Editor of IEEE Trans. Medical Imaging (TMI) (recognized as "Outstanding Associate Editor" by TMI), Medical Physics, and Machine Learning Science and Technology. He is Fellow of IEEE, SPIE, OSA, AIMBE, AAPM, and AAAS.